\documentstyle[12pt,a4,rotate,epsf]{article}

\newcommand{\qpm}{q_{\pm}}
\newcommand{\qmp}{q_{\mp}}
\newcommand{\nn}{\nonumber}
\newcommand{\Slash}{\not\!}
\newcommand{\al}{\alpha_{s}}
\newcommand{\lnp}{\big<\!\!:\!}
\newcommand{\rnp}{\!:\!\!\big>}
\newcommand{\Cbar}{\overline C}

\newcommand{\smvs}{\vbox{\vskip 8mm}}
\newcommand{\bmvs}{\vbox{\vskip 10mm}}
\newcommand{\qq}{\lnp\bar qq\rnp^{(0)}}
\newcommand{\qFq}{\lnp g_{s}\bar q\sigma Fq\rnp^{(0)}}

\newcommand{\lnpas}{\big<\frac{\alpha_{s}}{\pi}\!:\!}

\newcommand{\newsection}[1]{\section{#1}\setcounter{equation}{0}}

\begin{document}

\bibliographystyle{physics}


\begin{titlepage}
\begin{flushright}
{\sf\large TUM-T31-21/92}\\
{\sf\large July 1992}
\end{flushright}
\vspace{10mm}
\begin{center}
{\huge Current correlators to all orders in\\
\vspace*{4mm}
the quark masses}
{\Large \footnote{Supported by the German Bundesministerium f\"ur Forschung
und Technologie, under the contract 06 TM 761.} }
\vspace{15mm}\\
{\normalsize Matthias JAMIN${}^{2}$, Manfred M\"UNZ} \\
\ \\
{\small\sl Physik Department, Technische Universit\"at M\"unchen,
 D-8046 Garching, FRG.}\\
\vspace{-2mm}
{\small\sl Email: \phantom{m}jamin @ feynman.t30.physik.tu-muenchen.de}\\
\vspace{-2mm}
{\small\sl \hspace{10mm} mmuenz @ feynman.t30.physik.tu-muenchen.de}\\
{\small\sl ${}^{2}$ Address after October, 1st: Division TH, CERN,
                    CH-1211 Geneva 23.}
\vspace{15mm}\\
{\bf Abstract}
\end{center}
\noindent
The contributions to the coefficient functions of the quark and the
mixed quark-gluon condensate to mesonic correlators are calculated
for the first time
to all orders in the quark masses, and to lowest order in the strong
coupling constant. Existing results on the coefficient functions of
the unit operator and the gluon condensate are reviewed. The proper
factorization of short- and long-distance contributions in the
operator product expansion is discussed in detail. It is found that to
accomplish this task rigorously the operator product expansion has to be
performed in terms of {\em non}-normal-ordered condensates. The resulting
coefficient functions are improved with the help of the renormalization
group. The scale invariant combination of dimension 5 operators,
including mixing with the mass operator, which is needed for the
renormalization group improvement, is calculated in the leading order.
\end{titlepage}

\newpage
\setcounter{page}{1}


\newsection{Introduction}

Since the pioneering papers by Shifman, Vainshtein, and Zakharov (SVZ)
\cite{svz:79}, QCD sum rules have played a major role in extracting
parameters describing the structure of the QCD vacuum, the so called
QCD condensates, as well as for
calculating hadronic parameters, e.g. masses and decay constants of
mesons and baryons from the fundamental QCD Lagrangian. This procedure
assumes the ``duality'' of a hadronic versus a partonic description,
the partons here being quarks and gluons.

The calculation of the partonic part is performed in the framework of
the operator product expansion (OPE) \cite{wil:69}, most commonly
including operators up to dimension 6, whose contributions signal the
breakdown of perturbation theory for low momenta and parameterize
non-perturbative effects. A large amount of results on the coefficient
functions in the OPE exists in the literature, usually invoking some
additional approximation, as for example small or large quark masses,
or equal masses in the case of mesonic correlation functions. For further
information the reader should consult refs. \cite{svz:79} and
\cite{bec:81}--\nocite{rry:85,bag:86,nar:89,gen:90a,gen:90b}\cite{bag:92}.

An important aspect which has to be mentioned in this context is the
proper factorization of short- and long-distance contributions in the
OPE. This means that the coefficient functions describing the
short-distance part of a given correlation function should be free
from dependences on the infrared structure of the theory, e.g. mass
logarithms, and all such dependences should be absorbed into the
corresponding matrix elements of operators, the condensates, which shall
contain the long-distance contributions. This ``cancellation of mass log's''
has been already discussed in the literature \cite{gen:84,bro:84,bro:85}.
We shall deal further with this point below.

In this work we present new results on the coefficient functions
for the quark and the mixed quark-gluon condensate in the unequal
mass case, to all orders in the quark masses, and to lowest order in
perturbation theory, for scalar, pseudoscalar, vector, and axialvector
mesonic correlators. These results are obtained without any of the
approximations mentioned above. In addition, for completeness we review
the corresponding results for the unit operator and the gluon condensate.
Extensive comparison of our results with results existing in the
literature is made in various limiting cases. We explicitly show that
all mass logarithms, which for small masses correspond to long-distance
contributions, can be absorbed into the QCD condensates only,
if the OPE is performed in terms of {\em non}-normal-ordered operators.
This observation was already made in \cite{che:85,spi:88}. For an
application see also ref.~\cite{bra:92}. The resulting
coefficient functions are improved with the help of the renormalization
group. In this context, the scale invariant combination of dimension 5,
including mixing with the mass operator, is calculated.

An application of our results to an improved determination of the current
strange quark mass, as well as a discussion of higher order $\alpha_{s}$
corrections, will be presented in a forthcoming publication \cite{jam:92}.

In sect.~2, we summarize some known facts on the OPE. The new results
for the coefficient functions of the quark and mixed condensate are
presented in sects.~4 and 6, and the corresponding results for the
unit operator and the gluon condensate are given in sects.~3 and 5.
The proper factorization in the OPE and the cancellation of mass log's
is discussed in sect.~7, and in sect.~8 we show how to improve the
coefficient functions with the help of the renormalization group (RG).
Finally, sect.~9 summarizes our results.

\newsection{The Coefficient Functions}

Let us first define the relevant two-point and coefficient functions
which we are going to examine, and then discuss some of their general
properties in the framework of the OPE
\cite{wil:69,zim:73}. The scalar, pseudoscalar, vector, and axialvector
two-point functions are defined to be the matrix elements,
between the physical vacuum $\vert\,\Omega\big>$, of the time-ordered
product of the corresponding currents,
\begin{equation}
\Pi^{\Gamma}(q) \; \equiv \; i \, \big<\Omega\,\vert \, TN_{a}\{\,
\widetilde j_{\Gamma}(q) \, j_{\Gamma}^{\dagger}(0)\}\vert\,\Omega\big> \, ,
\label{eq:2.1}
\end{equation}
where $\Gamma$ stands for one of the Dirac-matrices $\Gamma\in\{1,i\gamma_{5},
\gamma_{\mu},\gamma_{5}\gamma_{\mu}\}$, specifying the quantum numbers
of the current (S, P, V, A respectively), the tilde sign will always
denote quantities Fourier-transformed to momentum space, and $N_{a}$
is a suitable renormalization procedure for the operator product
\cite{zim:73,col:84,sho:91}. Throughout this work we will use dimensional
regularization in $D=4-2\,\varepsilon$ dimensions \cite{tho:72,lei:75}
and the modified minimal subtraction ($\overline{MS}$) scheme \cite{bar:78}.

The current $j_{\Gamma}$ shall take the form
\begin{equation}
j_{\Gamma}(x) \; \equiv \; :\!\bar Q(x)\Gamma q(x)\!: \, ,
\label{eq:2.2}
\end{equation}
where $Q$ and $q$ are quarks of possibly different flavour with masses
$M$ and $m$ respectively. If $M$ and $m$ differ, we shall always assume
$M>m$. To the order we are working, the normal-ordering
is sufficient to renormalize the current. However, in higher orders,
additional subtractions are required.

Using the Ward-identity relating the divergence of the vector
(axialvector) two-point function to the scalar (pseudoscalar) two-point
function \cite{bec:81,bro:81}, we can express $\Pi^{\Gamma}(q)$ in terms
of 4 Lorentz-scalar two-point functions $\Pi^{I}(q^{2})$, with
$I\in\{S,P,V,A\}$. Then the vector (axialvector) two-point function
can be written as
\begin{eqnarray}
\Pi_{\mu\nu}^{V,A}(q) & = & \Big[\, \frac{q_{\mu}q_{\nu}}{q^{2}} -
g_{\mu\nu} \,\Big] \; \Pi^{V,A}(q^{2}) + g_{\mu\nu} \, \frac{
\big(M\mp m\big)^{2}}{q^{2}} \, \Pi^{S,P}(q^{2}) \nn \\
\smvs
& \phantom{=} & + \,g_{\mu\nu} \, \frac{\big(M\mp m\big)}{q^{2}} \,
\Big[\, \big<\bar QQ\big>^{(1)} \mp \big<\bar qq\big>^{(1)} \,\Big] \, ,
\label{eq:2.3}
\end{eqnarray}
with $\Pi^{S,P}(q^{2})$ being the scalar (pseudoscalar) two-point function.
In the following, $\big<\ldots\big>^{(1)}$ will always denote matrix
elements between the physical vacuum $\vert\,\Omega\big>$ (condensates),
the superscript indicating renormalization to the order considered (1-loop
in this case). Please note that the quark condensates in eq. (\ref{eq:2.3})
are {\em not} normal-ordered. The reason for this will be thoroughly
discussed below. On the other hand, because of this fact, the third term
in eq. (\ref{eq:2.3}) depends on the renormalization scale
--- $m\big<\bar qq\big>$ is no longer renormalization group invariant
\cite{spi:88} --- but this dependence is cancelled by the second term.

In the framework of the OPE, $\Pi^{I}(q^{2})$ can be expanded in terms
of local operators $O_{i}$ times Wilson-coefficient functions $C_{i}$,
$C_{1}$ denoting the coefficient function of the unit operator,
\begin{equation}
\Pi^{I}(q^{2}) \; = \; \Cbar_{1}^{\,I}\big(q^{2},M,m,\mu^{2},\alpha_{s}\big)
+ \sum_{i}\,\frac{\Cbar_{i}^{\,I}\big(q^{2},M,m,\mu^{2},\alpha_{s}
\big)}{(q^{2})^{[(n_{i}-1)/2]}} \, \lnp O_{i}(0) \rnp^{(0)} \, ,
\label{eq:2.4}
\end{equation}
thereby separating the short-distance dynamics, described by the
Wilson-coefficient functions, from the long-distance behaviour,
included in the operators $O_{i}$. The bar indicates coefficient functions
corresponding to tree-level matrix elements, $\mu$ is a renormalization
scale in the $\overline{MS}$-scheme, and the $n_{i}$ are the canonical
dimensions of the $O_{i}$. With $[(n_{i}-1)/2]$ denoting the integer part
of $(n_{i}-1)/2$, it follows that the $\Cbar_{i}^{\,I}$ have dimension~0
or 1, depending on the dimension of $O_{i}$ being even or odd.

Proper factorization of short- and long-distance contributions in the
OPE requires the calculation of the matrix elements in eq. (\ref{eq:2.4})
at the same order as the coefficient functions. This procedure is well
known from deep inelastic scattering \cite{bar:78,flo:78} and weak
decays \cite{bur:90}. In this way, taking into account mixing of the
operators under renormalization, all dependences on the infrared
structure of the theory are absorbed into the matrix elements.
Calculating the $\Cbar_{i}^{\,I}$ straightforwardly using Wick's theorem,
naturally leads to normal-ordered operators. However, as has been proven
by \cite{che:82,tka:83,lle:88}, in minimal subtraction schemes the
coefficient functions are analytic in the masses, if the OPE
is performed in terms of {\em non}-normal-ordered operators. This feature
is in general not present in non-minimal schemes. Since normal-ordering
is a non-minimal scheme, using normal-ordered operators there appear
terms of the form $m^{a}\ln^{b}m^{2}/\mu^{2}$ in the coefficient function
$\Cbar_{1}$. These terms would make it impossible to calculate non-leading
contributions, because, while summing up the usual perturbative
logarithms by setting $\mu^{2}=-q^{2}$, we would acquire possibly large
log's of the form $\ln\,(m^{2}/\!-\!q^{2})$. For small masses they are remnants
of the long-distance structure of the theory and should be absorbed into
the corresponding condensates. Therefore, to arrive at physically sensible
coefficient functions, we have to express the tree-level matrix elements
$\lnp O_{i} \rnp^{(0)}$ in terms of the renormalized $\big<O_{i}\big>^{(1)}$,
and, working with {\em non}-normal-ordered operators, also mixing with the
unit operator under renormalization has to be taken into account.
This will be done below.

In our analysis we shall include operators up to dimension~5, namely
\begin{equation}
O_{i} \; \in \; \Big\{\, \big(\bar QQ\big),\, \big(\bar qq\big),\,
\big(\frac{\alpha_{s}}{\pi}FF\big),\, \big(g_{s}\bar Q\sigma FQ\big),\,
\big(g_{s}\bar q\sigma Fq\big) \,\Big\} \, ,
\label{eq:2.5}
\end{equation}
related to the quark condensates,
the gluon condensate and the mixed condensates
having dimension 3, 4, and 5 respectively. Implicitly, summation over
colour, spinor, and Lorentz indices is assumed, and the $\sigma$-matrix
is defined as $\sigma_{\mu\nu}=i/2\,[\gamma_{\mu},\gamma_{\nu}]$.
Coefficient functions in various approximations up to dimension~8
have been calculated, and can be found for example in refs. \cite{svz:79},
\cite{bec:81}--\nocite{rry:85,bag:86,nar:89,gen:90a}\cite{gen:90b}, and
\cite{bro:85,nik:83,bag:85}.

Calculating the diagrams of fig.~1, in the leading order one obtains
the following relations \cite{gen:84,spi:88}
\begin{eqnarray}
\big<\bar QQ(\mu)\big>^{(1)} & = & \lnp\bar QQ\rnp^{(0)} - \frac{N}
{4\pi^{2}}\,M^{3} \Big[ \ln \frac{M^2}{\mu^{2}}-1 \Big]  -
\frac{1}{12 M} \, \lnpas FF\rnp^{(0)} \; , \nn \\
\bmvs
\big<\frac{\alpha_{s}}{\pi}FF\big>^{(1)} & = & \lnpas FF\rnp^{(0)} \, ,
\label{eq:2.6} \\
\bmvs
\big<g_{s}\bar Q\sigma FQ(\mu)\big>^{(1)} & = & \lnp g_{s}\bar Q\sigma
FQ\rnp^{(0)} + \frac{M}{2} \ln \frac{M^2}{\mu^{2}}\,\lnpas FF\rnp^{(0)}\, ,\nn
\end{eqnarray}
where $N$ is the number of colours. The corresponding relations for
the $q$-quark condensates are obtained through the replacements
$Q\rightarrow q$ and $M\rightarrow m$. Since the gluon condensate is
already ${\cal O}(\alpha_{s})$, it does not get renormalized to the
order we are working.

Inserting these relations into eq. (\ref{eq:2.4}), we find the infrared
stable coefficient functions $C_{i}^{I}$,
\begin{eqnarray}
C_{1}^{I} & = & \Cbar_{1}^{\,I} + \frac{N}{4\pi^{2}} \biggl\{\,
\frac{M^{3}}{q^{2}}\Big[\ln\frac{M^2}{\mu^{2}}-1\Big]\,\Cbar_{\bar QQ}^{\,I}
+ \frac{m^{3}}{q^{2}}\Big[\ln\frac{m^2}{\mu^{2}}-1\Big]\,\Cbar_{\bar qq}^{\,I}
\,\biggr\} \; , \label{eq:2.7} \\
\smvs
C_{\bar QQ}^{I} & = & \Cbar_{\bar QQ}^{\,I} \phantom{_{F}}
\qquad\qquad \hbox{and} \qquad\qquad \phantom{_{F}}
C_{\bar qq}^{I} \; = \; \Cbar_{\bar qq}^{\,I} \; , \label{eq:2.8} \\
\bmvs
C_{FF}^{I} & = & \Cbar_{FF}^{\,I} + \frac{1}{12M}\,\Cbar_{\bar QQ}^{\,I} +
\frac{1}{12\,m}\,\Cbar_{\bar qq}^{\,I}
- \frac{M}{2q^{2}}\ln\frac{M^2}{\mu^{2}}\,\Cbar_{\bar QFQ}^{\,I}
- \frac{m}{2q^{2}}\ln\frac{m^2}{\mu^{2}}\,\Cbar_{\bar qFq}^{\,I} \, ,
\nn \\ & & \label{eq:2.9} \\
C_{\bar QFQ}^{I} & = & \Cbar_{\bar QFQ}^{\,I}
\qquad\qquad \hbox{and} \qquad\qquad
C_{\bar qFq}^{I} \; = \; \Cbar_{\bar qFq}^{\,I} \, . \label{eq:2.10}
\end{eqnarray}
In terms of these coefficient functions, the OPE takes the form
\begin{equation}
\Pi^{I}(q^{2}) \; = \; C_{1}^{I}\big(q^{2},M,m,\mu^{2},\alpha_{s}\big)
+ \sum_{i}\,\frac{C_{i}^{I}\big(q^{2},M,m,\mu^{2},\alpha_{s}
\big)}{(q^{2})^{[(n_{i}-1)/2]}} \, \big< O_{i}(\mu) \big>^{(1)} \, .
\label{eq:2.11}
\end{equation}

In the following sections we calculate the coefficient functions of
eqs. (\ref{eq:2.7})--(\ref{eq:2.10}) for the unit operator and the
quark, gluon, and mixed condensate to all orders in the quark masses,
and show explicitly that
up to operators of dimension~7, they are analytic functions of the
masses. Of course, as we shall also demonstrate, the coefficient
functions may still contain non-analytic pieces of higher dimension,
which are only cancelled through mixing with higher dimensional operators.
In our example this is the case for $C_{1}^{I}$ and $C_{FF}^{I}$.
This point will be discussed further in section~7.

\newsection{The Perturbative Coefficient}

The perturbative coefficient function for two different quark masses
has been already calculated in ref. \cite{gen:90b}, in the leading
order, as well as to ${\cal O}(\alpha_{s})$. For completeness, and
for further reference, we give here our results for the coefficient
functions $\Cbar_{1}^{\,I}$, which are in agreement with \cite{gen:90b},
however using a slightly more compact notation.
\begin{eqnarray}
\Cbar_{1}^{\,S,P}(q^{2}) & = & -\, \frac{N}{8\pi^{2}} \, \biggl\{\,
\qpm^{2}\, I_S\big(q^{2},M,m,\mu^{2}\big)-M^{2}\, l_{M}-m^{2}\, l_{m}
\,\biggr\} \,, \label{3.1} \\
\bmvs
\Cbar_{1}^{\,V,A}(q^{2}) & = & -\, \frac{N}{12\pi^{2}} \, \biggl\{\,
\biggl[\, q^{2}+M^{2}+m^{2}-2\,\frac{(M^{2}-m^{2})^{2}}{q^{2}} \,\biggr]\,
I_S\big(q^{2},M,m,\mu^{2}\big) \nn \\
\smvs
-M^{2}\, l_{M}&&\hspace{-1cm}-\,m^{2}\, l_{m}+2\,\frac{(M^{2}-m^{2})}
{q^{2}}\,\Big[\,
M^{2}\, l_{M}-m^{2}\, l_{m}\,\Big]+\frac{q^{2}}{3}-M^{2}-m^{2} \,\biggr\} \,.
\label{3.2}
\end{eqnarray}
Here $l_{M}\equiv\ln(M^{2}/\mu^{2})-1$, and $I_S(q^{2},M,m,\mu^{2})$ is
the scalar one-loop integral,
\begin{eqnarray}
I_S\big(q^{2},M,m,\mu^{2}\big) & \equiv & \int_{0}^{1} dx\,\ln\frac{xM^{2}+
(1-x)\,m^{2}-x(1-x)\,q^{2}}{\mu^{2}} \nn \\
\bmvs
& = & \frac{u\,q_{-}^{2}}{q^{2}}\,\ln\frac{u+1}{u-1} + \frac{(M^{2}-m^{2})}
{q^{2}}\,\ln\frac{M}{m}+\ln\frac{Mm}{\mu^{2}}-2 \,, \quad\label{3.3}
\end{eqnarray}
with
\begin{equation}
\qpm^{2} \; \equiv \; q^{2} - \big(M\pm m\big)^{2}
\quad\qquad \hbox{and} \quad\qquad
u \; \equiv \; \sqrt{1-\frac{4Mm}{q_{-}^{2}}} \, .
\label{3.4}
\end{equation}

To show explicitly the appearance of mass logarithms in the coefficient
functions, we also present their expansions in the quark masses up to
order $m^{4}$.
\begin{eqnarray}
\Cbar_{1}^{\,S,P}(q^{2}) & = & \frac{N}{8\pi^{2}} \, \biggl\{\,
2\,\qpm^{2}-\big(\qpm^{2}-M^{2}-m^{2}\big)\,\ln\frac{-q^{2}}{\mu^{2}} \nn \\
\smvs
& & -\,\Big[\,\frac{3}{2}\,M^{4}\pm2M^{3}m+2M^{2}m^{2}\pm2Mm^{3}
+\frac{3}{2}\,m^{4}\,\Big]\,\frac{1}{q^{2}} \nn \\
\smvs
& & +\,\Big[\,(M\pm2m)\,M^{3}\ln\frac{M^{2}}{-q^{2}}+(m\pm2M)\,
m^{3}\ln\frac{m^{2}}{-q^{2}}\,\Big]\,\frac{1}{q^{2}} \,\biggr\}
\,, \label{3.5} \quad \\
\bmvs
\Cbar_{1}^{\,V,A}(q^{2}) & = & \frac{N}{12\pi^{2}} \, \biggl\{\,
\frac{5}{3}\,q^{2}+3\,\big(M^{2}+m^{2}\big)-q^{2}\,\ln\frac{-q^{2}}{\mu^{2}}
\nn \\
\smvs
& & \hspace{-1cm}-\,3\,\Big[\,\frac{1}{2}\,M^{4}-2M^{2}m^{2}+
\frac{1}{2}\,m^{4}+M^{4}\,\ln\frac{M^{2}}{-q^{2}}+m^{4}\,\ln\frac{m^{2}}
{-q^{2}}\,\Big]\,\frac{1}{q^{2}} \,\biggr\} \,. \label{3.6} \quad
\end{eqnarray}
As has been discussed in the previous section, these mass logarithms are
cancelled after inclusion of the mixing with the quark condensate. This
will be performed in section~7. Expressions for the $\Cbar_{1}^{\,I}$,
only expanding in $m$, are provided in appendix~A.

\newsection{The Quark Condensate}

The calculation of the coefficient functions for the quark condensate
follows closely the method used in ref. \cite{ynd:89}. The contribution
of the quark condensate to the two-point function $\Pi^{\Gamma}(q)$,
shown in fig.~2, is given by
\begin{equation}
\Pi^{\Gamma}_{\bar qq+\bar QQ}(q) \; = \; - \int\!\frac{d^{D}\!p}{(2\pi)^D}
\, \lnp \bar q(0)\Gamma S_{Q}(p-q)\Gamma\tilde q(p)\rnp^{(0)} + \;
(q\leftrightarrow Q) \, .
\label{eq:4.1}
\end{equation}
Here $S_{Q}(p-q)=(\Slash p\,-\!\Slash q-M)^{-1}$ is the free quark propagator.

A necessary ingredient for calculating the coefficient functions to all
orders in the quark masses is a closed expression for the non-local quark
condensate. In $x$-space this expression reads \cite{bag:86,ynd:89,eli:88}
\begin{equation}
\lnp\bar q_{\alpha}(0)q_{\beta}(x)\rnp^{(0)}_{\bar qq} \; = \; \frac{1}{4m}\,
\qq\,\Gamma\left(\frac{D}{2}\right)(i\!\Slash\partial+m)_{\beta\alpha}
\sum_{n=0}^{\infty} \frac{(-m^{2}x^{2}/4)^{n}}{n!\,\Gamma(n+D/2)} \, ,
\label{eq:4.2}
\end{equation}
where the index $\bar qq$ denotes the projection onto the local quark
condensate. One should remark that the non-local quark condensate
has, of course, additional
contributions from higher dimensional operators (see section~6). The
sum in (\ref{eq:4.2}) can be expressed in terms of Bessel functions,
\begin{equation}
\sum_{n=0}^{\infty} \frac{(-m^{2}x^{2}/4)^{n}}{n!\,\Gamma(n+D/2)} \; = \;
\left(\frac{2}{\sqrt{m^{2}x^{2}}}\right)^{D/2-1} J_{D/2-1}\big(
\sqrt{m^{2}x^{2}}\big) \, ,
\label{eq:4.3}
\end{equation}
but like in \cite{ynd:89}, we prefer to work explicitly with the expanded
form. Since we shall derive similar relations for the mixed condensate,
we skip the derivation of eq. (\ref{eq:4.2}), and refer the reader to
appendix~A of ref. \cite{eli:88}.

The corresponding $p$-space expression for the non-local quark condensate
is given by \cite{ynd:89}
\begin{equation}
\lnp\bar q_{\alpha}(0)\tilde q_{\beta}(p)\rnp^{(0)}_{\bar qq} \, = \,
\frac{(2\pi)^{D}}{4m}\, \qq \,\Gamma\left(\frac{D}{2}\right)
(\Slash p+m)_{\beta\alpha} \sum_{n=0}^{\infty} \frac{(m^{2}/4\,
\partial_{p}^{\,2})^{n}}{n!\,\Gamma(n+D/2)}\, \delta^{D}(p) \, .
\label{eq:4.4}
\end{equation}
It is easy to verify that the quark condensate (\ref{eq:4.4}) satisfies a
free equation of motion,
\begin{equation}
\big(\!\Slash p-m\big) \, \lnp\bar q(0)\tilde q(p)\rnp^{(0)}_{\bar qq}
\; = \; 0 \, .
\label{eq:4.5}
\end{equation}
This is very useful, as it allows to replace an arbitrary non-singular
function $f(p^{2},p,\ldots)$ by $f(m^{2},p,\ldots)$ in integrals of the type
\begin{equation}
\int d^{D}\!p \, f(p^{2},p,\ldots) \, \lnp\bar q(0)\tilde q(p)
\rnp^{(0)}_{\bar qq} \, ,
\label{eq:4.6}
\end{equation}
thereby greatly simplifying the calculation.

After performing the momentum integration and using the relations
\begin{equation}
\left[\frac{m^{2}}{4}\,\partial_{p}^{\,2}\right]^{n} \frac{1}
{[\,q^{2}-2q\cdot p-M^{2}+m^{2}\,]} \; = \; \frac{(2n)!\,(m^{2}q^{2})^{n}}
{[\,q^{2}-2q\cdot p-M^{2}+m^{2}\,]^{2n+1}} \, ,
\label{eq:4.7}
\end{equation}

\begin{eqnarray}
f(z) & \equiv & \Gamma(D/2) \sum_{n=0}^{\infty}
\frac{(2n)!}{n!\,\Gamma(n+D/2)} \, z^{n} \nn \\
\smvs
& = & _{2}F_{1}(1,1/2;D/2;4z) \; \stackrel{D=4}{=} \;
\frac{1}{2z} \, \Big[\, 1 - \sqrt{1-4z} \,\Big] \, ,
\label{eq:4.8}
\end{eqnarray}
where $_{2}F_{1}$ is the Hypergeometric function \cite{abr,gra},
we arrive at the following final results for the two-point functions
in $D=4$ dimensions:
\begin{eqnarray}
\Pi_{\bar qq}^{S,P}(q^{2}) & = & - \, \frac{\qq}{2\,m} \,
\left\{\, 1-\frac{\qpm^{2}}{[\,q^{2}-M^{2}+m^{2}\,]} \,
f(z_{m})\, \right\} \, , \label{eq:4.9} \\
\bmvs
\Pi_{\bar qq}^{V,A}(q^{2}) & = & - \, \frac{\qq}{3\,m} \,
\biggl\{\, 1+2\,\frac{(M^{2}-m^{2})}{q^{2}} \nn \\
\smvs
& \phantom{=} & - \,\frac{[\,q^{2}+M^{2}+m^{2}-2\,(M^{2}-m^{2})^{2}
/q^{2}\,]}{[\,q^{2}-M^{2}+m^{2}\,]} \, f(z_{m})\, \biggr\} \, , \qquad
\label{eq:4.10}
\end{eqnarray}
with
\begin{equation}
z_{m} \; \equiv \; \frac{m^{2}q^{2}}{[\,q^{2}-M^{2}+m^{2}\,]^{2}} \, .
\label{eq:4.11}
\end{equation}
The corresponding functions for the heavy quark $Q$,
$\Pi_{\bar QQ}^{I}(q^{2})$, can be obtained through the replacements
$q\rightarrow Q$ and $m\leftrightarrow M$. In the equal mass case,
$M=m$, our results agree with the various results given by the authors
of refs. \cite{bag:86,ynd:89,ste:89}, except for the axial current in
\cite{bag:86}.

To examine the structure of the coefficient functions more explicitly,
and to be able to compare with other previous results, we expand the
two-point functions in the quark masses up to operators of dimension~6.
For the coefficient functions of the quark condensate this leads to
\begin{eqnarray}
C_{\bar qq}^{S,P}(q^{2}) & = & \Cbar_{\bar qq}^{\,S,P}(q^{2}) \; = \;
\phantom{m\,} \biggl[\, \mp M-\frac{m}{2}\mp\frac{M^{3}}{q^{2}} \;\biggr] \, ,
\label{eq:4.12} \\
\bmvs
C_{\bar qq}^{V,A}(q^{2}) & = & \Cbar_{\bar qq}^{\,V,A}(q^{2}) \; = \;
m \,\biggl[\, 1+2\,\Big(M^{2}-\frac{m^{2}}{3}\Big)\frac{1}{q^{2}} \,\biggr]\,,
\label{eq:4.13}
\end{eqnarray}
and related expressions for the $Q$-quark. Again, in appendix~A we provide
results for solely expanding in $m$ up to order $m^{3}$. These results
also agree with the results given by Generalis \cite{gen:90a}.

\newsection{The Gluon Condensate}

The contribution of the gluon condensate to all orders in the quark masses
has already been presented in refs. \cite{rry:85,gen:90a}. We agree with
these results and, for completeness, and further reference, cite here the
corresponding expressions.
\begin{eqnarray}
\Pi_{FF}^{S,P}(q^{2}) & = & \frac{-1}{48}\,\frac{q^{2}}{\qmp^{4}}\,
\lnpas FF\rnp^{(0)} \biggl\{\, \frac{3(3+u^{2})(1-u^{2})}{2u^{3}} \log
\frac{u+1}{u-1} - \frac{3u^{4}+4u^{2}+9}{u^{2}(1-u^{2})} \,\biggr\} \nn \\
\smvs
& & \mp \; \frac{1}{12Mm}\,\lnpas FF\rnp^{(0)} \, , \label{eq:5.1} \\
\bmvs
\Pi_{FF}^{V,A}(q^{2}) & = & \frac{1}{48}\,\frac{q^{2}}{\qmp^{4}}\,
\lnpas FF\rnp^{(0)} \biggl\{\, \frac{3(1+u^{2})(1-u^{2})^{2}}{2u^{5}} \log
\frac{u+1}{u-1} - \frac{3u^{4}-2u^{2}+3}{u^{4}} \,\biggr\} \nn \\
& & \label{eq:5.2}
\end{eqnarray}
For $P$ and $A$ an additional change $M\rightarrow-M$ in $u$, being
equivalent to $u\rightarrow1/u$, has to be performed. In the equal mass
case our results for $\Pi_{FF}^{S,P}$ agree with those of Bag{\'a}n et.al.
\cite{bag:86}.

We would like to point out that the non-transverse part for the
two-point function $\Pi_{FF,\,\mu\nu}^{V,A}$ of ref. \cite{rry:85}
(eq. 3.53), and our scalar (pseudoscalar) two-point function
$\Pi_{FF}^{S,P}$, differ by the $q$-independent piece in eq.
(\ref{eq:5.1}). Writing eq.~(\ref{eq:2.3}) in terms of tree-level
condensates would cancel this additional term, but then the Ward-identity
would no longer be valid.

Upon expansion in the quark masses up to operators of dimension~6,
we find for the coefficient functions $\Cbar_{FF}^{\,I}$
\begin{eqnarray}
\Cbar_{FF}^{\,S,P} & = & - \,\frac{1}{24} \pm \frac{1}{12}\Big(
\frac{M}{m} + \frac{m}{M} \Big) - \frac{1}{12q^{2}}\big(M^{2}+m^{2}\big)\nn \\
\smvs
& & \pm \,\frac{1}{12q^{2}}\Big(\frac{M^{3}}{m}+\frac{m^{3}}{M}\Big)
\pm \frac{Mm}{4q^{2}}\Big[\, 3+2\log \frac{Mm}{-q^{2}} \,\Big] \, ,
\label{eq:5.4} \\
\bmvs
\Cbar_{FF}^{\,V,A} & = & - \,\frac{1}{12} - \frac{1}{6q^{2}}
\big(M^{2}+m^{2}\big) \, . \label{eq:5.5}
\end{eqnarray}
Like in the case of the quark condensate, in appendix~A we provide
results for the $\Cbar_{FF}^{\,I}$, expanded only in $m$.

A few remarks are in order here: as is evident, at this intermediate
stage there appear $1/m$ as well as $m^{2}\log m$ terms. Nevertheless,
they are remnants of the long-distance structure of the vacuum condensates
and will cancel in the final result for the coefficient functions
$C_{FF}^{I}$, once the additional contributions through mixing
(\ref{eq:2.9}) have been included. This cancellation of mass singularities
has already been extensively discussed in \cite{gen:84,bro:84,bro:85}.
We present the coefficients $C_{FF}^{I}$ in section~7.

\newsection{The Mixed Condensate}

The calculation of the contribution to the mixed condensate to all orders
in the masses is somewhat more complicated than in the case of the quark
condensate. We shall therefore discuss its evaluation in slightly more
detail. The two diagrams contributing to the mixed condensate are shown
in fig.~3.

Let us first consider diagram 3a. Working in the coordinate gauge, this
graph also arises from eq. (\ref{eq:4.1}), since in the interacting case
$q(x)$ has to be expanded in terms of covariant derivatives (see for
example \cite{pas:84}), and hence yields contributions which involve
gluon fields. The procedure for the calculation of the mixed condensate
contribution is similar to the one presented in the appendices~A and C of
ref. \cite{eli:88}. For the explicit steps, the reader is referred to this
publication. Our result for the non-local quark condensate contribution
to the mixed condensate is
\begin{eqnarray}
\lnp\bar q_{\alpha}(0)q_{\beta}(x)\rnp^{(0)}_{\bar qFq} & = &
-\,\frac{1}{8m^{3}}\,\qFq\,\Gamma\left(\frac{D}{2}\right) \,\cdot \nn \\
\smvs
& & \cdot\,\sum_{n=0}^{\infty} \Big[\, (n-1)i\!\Slash\partial+n\,m\,\Big]
_{\beta\alpha}\,\frac{(-m^{2}x^{2}/4)^{n}}{n!\,\Gamma(n+D/2)} \, .
\label{eq:6.1}
\end{eqnarray}
Up to terms ${\cal O}(m^{3})$, the order calculated in ref. \cite{eli:88},
our result agrees with \cite{eli:88}.

Similar to the quark condensate, the mixed condensate contribution satisfies
an equation of motion. In $p$-space, this equation of motion reads
\begin{equation}
\big(p^{2}-m^{2}\big) \, \lnp\bar q(0)\tilde q(p)\rnp^{(0)}_{\bar qFq}
\; = \;-\,\frac{\qFq}{2\,\qq}\,\lnp\bar q(0)\tilde q(p)\rnp^{(0)}_{\bar qq} \,,
\label{eq:6.2}
\end{equation}
which also allows for simplification in momentum integrals. The replacement
in momentum integrals which can be made by means of eq. (\ref{eq:6.2}) has
the form
\begin{eqnarray}
\int d^{D}\!p \, f(p^{2},p,\ldots)\,\lnp\bar q(0)\tilde q(p)\rnp^{(0)}
_{\bar qFq} & \longrightarrow & \int d^{D}\!p \,\biggl\{\,f(m^{2},p,\ldots)
\,\lnp\bar q(0)\tilde q(p)\rnp^{(0)}_{\bar qFq} \nn \\
\smvs & & \hspace{-4.4cm}
- \,\frac{\qFq}{2\,\qq}\,\Big[\,\partial_{m^{2}}f(m^{2},p,\ldots)\,\Big]
\,\lnp\bar q(0)\tilde q(p)\rnp^{(0)}_{\bar qq} \,\biggr\} \,. \label{eq:6.2a}
\end{eqnarray}

The contribution from diagram 3b can be evaluated by considering the
insertion of one gluon field strength into the non-local quark condensate
of eq. (\ref{eq:4.2}) \cite{pas:84}. The corresponding expression for the
non-local mixed condensate is found to be
\begin{eqnarray}
\lnp g_{s}\bar q_{\alpha}(0)F_{\mu\nu}(0)\,q_{\beta}(x)\rnp^{(0)}_{\bar qFq}
& = & \frac{1}{4(D-1)(D-2)\,m^{2}}\,\qFq\,\Gamma\left(\frac{D}{2}\right)
\,\cdot \nn \\
\smvs & & \hspace{-3.75cm}
\cdot\, \Big[\,\big(\gamma_{\mu}\partial_{\nu}-\gamma_{\nu}\partial_{\mu}
\big)+m\,\sigma_{\mu\nu} \,\Big] \big(i\!\Slash\partial+m\big)_{\beta\alpha}
\sum_{n=0}^{\infty} \frac{(-m^{2}x^{2}/4)^{n}}{n!\,\Gamma(n+D/2)} \,. \quad
\label{eq:6.3}
\end{eqnarray}
Here $F_{\mu\nu}=t^{a}F_{\mu\nu}^{a}$, where $t^{a}$ are the generators of
the colour group $SU(N)$. The expansion of eq. (\ref{eq:6.3}) has been
calculated by the authors of ref. \cite{eli:88} up to order $m^{3}$, and
we agree with their result, except for the term ${\cal O}(m^{3})$.
The derivation of eq.~(\ref{eq:6.3}) is presented in appendix~B.

It is obvious from the structure of eq. (\ref{eq:6.3}), that the non-local
mixed condensate also satisfies the free equation of motion
\begin{equation}
\big(\!\Slash p-m\big) \, \lnp g_{s}\bar q_{\alpha}(0)F_{\mu\nu}(0)\,
q_{\beta}(x)\rnp^{(0)}_{\bar qFq} \; = \; 0 \, .
\label{eq:6.4}
\end{equation}

Using eqs. (\ref{eq:6.1}) and (\ref{eq:6.3}), as well as the corresponding
equations of motion, we obtain the following final results for the
contribution of the mixed condensate to the two-point function of eq.
(\ref{eq:2.1}):
\begin{eqnarray}
\Pi_{\bar qFq}^{S,P}(q^{2}) & = & - \, \frac{\qFq}{2\,m^{3}q_{\mp}^{2}}\,
\biggl\{\, q^{2}-M^{2}\pm Mm \nn \\
\smvs
& & - \,\frac{[\,(q^{2}-M^{2})^{2}\pm Mm(q^{2}-M^{2}+m^{2})-M^{2}m^{2}\,]}
{[\,q^{2}-M^{2}+m^{2}\,]} \, f(z_{m})\, \biggr\} \, , \quad\label{eq:6.5} \\
\bmvs
\Pi_{\bar qFq}^{V,A}(q^{2}) & = & - \, \frac{\qFq}{3\,m^{3}q^{2}q_{\pm}^{2}
q_{\mp}^{2}}\, \biggl\{\, [q^{2}+2M(M\mp m)](q^{2}-M^{2})^{2}-m^{2} q^{2}
(q^{2}+M^{2}) \nn \\
\smvs
& & - \,2Mm^{2}(M\mp m)(2M^{2}-m^{2})-\frac{P(q^{2},M,m)}{[\,q^{2}-M^{2}+
m^{2}\,]} \, f(z_{m})\, \biggr\} \, , \quad\label{eq:6.6}
\end{eqnarray}
with
\begin{eqnarray}
P(q^{2},M,m) & = & [q^{2}+2M(M\mp m)](q^{2}-M^{2})^{3} - M^{2}m^{2}q^{2}
(4M^{2}\mp6Mm+m^{2}) \nn \\
\smvs & - & \hspace{-2mm}
m^{2}q^{4}(q^{2}+M^{2}) + 2Mm^{2}(M\mp m)(3M^{4}-3M^{2}m^{2}+m^{4}) \,.
\label{eq:6.7}
\end{eqnarray}
As in the case of the quark condensate, the corresponding functions for
the heavy quark $Q$, $\Pi_{\bar QFQ}^{I}(q^{2})$, can be obtained through
the replacements $q\rightarrow Q$ and $m\leftrightarrow M$. In the
equal mass case, $M=m$, our results agree with the result given in
\cite{bag:86}. The expansion in one light mass $m$ up to operators of
dimension 6 for the vector current agrees with the result obtained in
\cite{gen:90a}.

Expanding the coefficient functions for the mixed condensate up to
operators of dimension 6 yields the well known expressions
\begin{equation}
C_{\bar qFq}^{S,P}(q^{2}) \; = \; \Cbar_{\bar qFq}^{\,S,P}(q^{2}) \; = \;
\pm\,\frac{M}{2} \, , \quad \hbox{and} \quad
C_{\bar qFq}^{V,A}(q^{2}) \; = \; \Cbar_{\bar qFq}^{\,V,A}(q^{2}) \; = \; 0 \,,
\label{eq:6.8}
\end{equation}
and related expressions for the $Q$-quark. Again, in appendix~A, we give
results for the expansion in one small mass $m$.

\newsection{The Cancellation of Mass Log's}

We are now in a position to calculate the infrared stable coefficient
functions $C_{1}^{I}$ and $C_{FF}^{I}$ of eqs. (\ref{eq:2.7}) and
(\ref{eq:2.9}). The result up to dimension 4 for the unit operator
is found to be
\begin{eqnarray}
C_{1}^{S,P}(q^{2}) & = & \frac{N}{16\pi^{2}}\, \biggl\{\, 4\,q_{\pm}^{2}-
\big(M^{4}+4M^{2}m^{2}+m^{4}\big)\frac{1}{q^{2}} \nn \\
\smvs & & \hspace{-2cm}
- \,2\Big[\, q_{\pm}^{2}-M^{2}-m^{2}+\big(M^{4}\pm2M^{3}m\pm2Mm^{3}+
m^{4}\big)\frac{1}{q^{2}} \,\Big]\ln\frac{-q^{2}}{\mu^{2}} \,\biggr\} \,,
\label{eq:7.1} \\
\bmvs
C_{1}^{V,A}(q^{2}) & = & \frac{N}{72\pi^{2}}\, \biggl\{\, 10\,q^{2}+
18\,\big(M^{2}+m^{2}\big)-9\,\big(\,3M^{4}-4M^{2}m^{2}+3m^{4}\big)
\frac{1}{q^{2}} \nn \\
\smvs
& & - \,6\Big[\,q^{2}-3\big(M^{4}+m^{4}\big)\frac{1}{q^{2}}\,\Big]
\ln\frac{-q^{2}}{\mu^{2}} \,\biggr\} \,, \label{eq:7.2} \\
\smvs
& & \hspace{-3.0cm}\hbox{and up to dimension 6 for the gluon condensate} \nn\\
\smvs
C_{FF}^{S,P}(q^{2}) & = & - \,\frac{1}{8}-\frac{(M^{2}+m^{2})}{12\,q^{2}}
\pm\frac{Mm}{4\,q^{2}}\,\Big[\,3-2\ln\frac{-q^{2}}{\mu^{2}}\,\Big] \,,
\label{eq:7.3} \\
\bmvs
C_{FF}^{V,A}(q^{2}) & = & \frac{1}{12}-\frac{(M^{2}+m^{2})}{18\,q^{2}} \,.
\label{eq:7.4}
\end{eqnarray}

It is evident that all mass logarithms have cancelled, and the resulting
coefficient functions are only polynomial in the quark masses. Only this
fact allows for a consistent summation of logarithmic corrections through
the choice $\mu^{2}=-q^{2}$. This task is accomplished in the next section.

However, the cancellation of mass log's only takes place up to the dimension
of operators, to which all contributions have been included consistently.
For example, the coefficient functions $C_{FF}^{I}(q^{2})$ still contain mass
log's of the form $Mm^{3}\ln m$, which were only cancelled, if mixing with
operators of dimension 8 had been included \cite{bro:85}.

The appropriate treatment of small masses in the OPE therefore is to
expand in the mass up to operators of the dimension in question. Analogously,
heavy masses have to be expanded in $1/M$. The case of heavy quark masses
will be discussed in a subsequent publication \cite{jam:93}.

\newsection{Renormalization Group Improved Coefficients}

The coefficient functions for the unit operator of eqs.
(\ref{eq:7.1}) and (\ref{eq:7.2}) satisfy an inhomogeneous
renormalization group equation,
\begin{equation}
\mu\,\frac{d}{d\mu}\,C_{1}^{I}\big(Q^{2}/\mu^{2}\big) \; = \; h_{0}^{I} \,,
\label{eq:8.1}
\end{equation}
where we have set $Q^{2}=-\,q^{2}$, and
\begin{eqnarray}
h_{0}^{S,P} & = & \frac{N}{4\pi^{2}}\,\Big[\, q_{\pm}^{2}-M^{2}-
m^{2}+\big(M^{4}\pm2M^{3}m\pm2Mm^{3}+m^{4}\big)\frac{1}{q^{2}} \,\Big] \,,
\label{eq:8.2} \\
h_{0}^{V,A} & = & \frac{N}{6\pi^{2}}\,\Big[\,q^{2}-3\big(M^{4}+m^{4}\big)
\frac{1}{q^{2}}\,\Big] \,. \label{eq:8.3}
\end{eqnarray}
The presence of the inhomogeneity
originates from the divergence of the current product of eq.~(\ref{eq:2.1})

The solution to this equation is given by
\begin{equation}
C_{1}^{I}\big(Q^{2}/\mu^{2}\big) \; = \; C_{1}^{I}\big(1\big) +
\frac{\pi h_{0}^{I}}{\beta_{0}}\left[\,\frac{1}{\al(Q^{2})}-
\frac{1}{\al(\mu^{2})}\,\right] \,.
\label{eq:8.4}
\end{equation}
This expression is the improved form for the coefficient function
$C_{1}^{I}$, where the leading $\ln(Q^{2}/\mu^{2})$ have been summed
up. $\beta_{0}$ is the leading coefficient in the expansion of the
$\beta$-function,
\begin{equation}
\mu\,\frac{d\al}{d\mu} \; = \; \al\beta(\al) \, , \qquad
\beta(\al) \; = \; \beta_{0}\,\frac{\al}{\pi}+\ldots \, , \qquad
\beta_{0} \; = \;-\,\frac{1}{6}(11N-2f) \,,
\label{eq:8.5}
\end{equation}
where $f$ denotes the number of quark flavours. Here and in the
following we have adopted the notation of Pascual and Tarrach
\cite{pas:84} for the renormalization group functions.

To improve on the contribution from higher dimensional operators,
we note, that the sum in eq.~(\ref{eq:2.11}) has to be independent
of $\mu$. Therefore, the choice $\mu^{2}=Q^{2}$ sums up the leading
log's, leaving us with condensates $\big<O_{i}(Q^{2})\big>$,
evaluated at the scale $Q^{2}$. These can now be expressed in terms
of condensates $\big<O_{i}(\mu_{0}^{2})\big>$, evaluated at some
fixed scale $\mu_{0}$, at which the numerical values of the condensates
are known, e.g. $\mu_{0}=1\,GeV$.

For simplicity, in the following we consider only one quark flavour
$q$, however keeping the $f$ dependence in the renormalization group
functions. The generalization to an arbitrary number of flavours should
be obvious.

In order to write the relation between $\big<O_{i}(Q^{2})\big>$ and
$\big<O_{i}(\mu_{0}^{2})\big>$, it is convenient to assemble the
operators $O_{i}$ into a vector $\vec O$,
\begin{equation}
\vec O^{T} \; = \; \Big(\,\big(g_{s}\bar q\sigma Fq\big),\,
\big(\frac{\alpha_{s}}{\pi}FF\big),\, \big(\bar qq\big),\, m\,\Big) \,,
\label{eq:8.6}
\end{equation}
in which we have included the mass as an operator. This allows us to
write the scale invariant combinations of the operators $O_{i}$ in
the form
\begin{equation}
\vec\phi^{\,T} \; = \;
\big(\, \phi_{3},\, \phi_{2},\, \phi_{1},\, \phi_{0} \,\big)^{T} \;
\equiv \; \hat R(\mu)\,\big<\vec O(\mu)\big> \,.
\label{eq:8.7}
\end{equation}
$\phi_{0}$ just is the invariant quark mass, the invariants $\phi_{1}$
and $\phi_{2}$ have been calculated in ref.~\cite{spi:88}, where also
next-to-leading order corrections can be found, and $\phi_{3}$ was
obtained in ref.~\cite{nar:83}, without taking into account mixing
with the mass operator.

The matrix $\hat R(\mu)$ is given by
\begin{equation}
\hat R(\mu) \; = \; \left(
\begin{array}{cccc}
\al^{d_{\sigma}^{\,0}}(\mu) & x\,\al^{d_{\sigma}^{\,0}-1}(\mu)\,m(\mu) &
y\,\al^{d_{\sigma}^{\,0}}(\mu)\,m^{2}(\mu) & z\,\al^{d_{\sigma}^{\,0}-1}
(\mu)\,m^{4}(\mu) \\
0& \pi\beta_{0}& 4\gamma_{m}^{0}\al(\mu)\,m(\mu) & \gamma_{0}^{0}\,m^{3}(\mu)\\
0 & 0 & m(\mu) & w\,\al^{-1}(\mu)\,m^{3}(\mu) \\
0 & 0 & 0 & \al^{d_{m}^{\,0}}(\mu) \\
\end{array} \right) \,,
\label{eq:8.8}
\end{equation}
where $\gamma_{m}^{0}$, $\gamma_{0}^{0}$, and $\gamma_{\sigma}^{0}$ are
the leading order anomalous dimensions of the mass operator, the vacuum
energy, and the mixed condensate, respectively,
\begin{equation}
\gamma_{m}^{0} \; = \; \frac{3(N^{2}-1)}{4N}\,, \qquad
\gamma_{0}^{0} \; = \; \frac{N}{2\pi}\,, \quad {\rm and} \quad
\gamma_{\sigma}^{0} \; = \; \frac{(N^{2}-5)}{4N}\,, \qquad
\label{eq:8.9}
\end{equation}
$d_{m}^{\,0}$ and $d_{\sigma}^{\,0}$ are defined to be the ratios
$d_{m}^{\,0}\equiv\gamma_{m}^{0}/\beta_{0}$ and
$d_{\sigma}^{\,0}\equiv\gamma_{\sigma}^{0}/\beta_{0}$, and finally,
$w$, $x$, $y$, and $z$ are given by
\pagebreak
\begin{eqnarray}
w & = & \frac{\gamma_{0}^{0}}{(\beta_{0}+4\gamma_{m}^{0})} \; = \;
\frac{3N^{2}}{\pi(7N^{2}+2Nf-18)} \,, \nn \\
x & = & \frac{-\pi}{(\beta_{0}+\gamma_{m}^{0}-\gamma_{\sigma}^{0})} \; = \;
\frac{6\pi N}{(8N^{2}-2Nf-3)} \,, \label{eq:8.10} \\
y & = & \frac{4\gamma_{m}^{0}(\beta_{0}+\gamma_{m}^{0}-\gamma_{\sigma}^{0}
-1)}{(\gamma_{\sigma}^{0}-\gamma_{m}^{0})(\beta_{0}+\gamma_{m}^{0}-
\gamma_{\sigma}^{0})} \; = \; \frac{-6(N^{2}-1)[\,8N^{2}+2N(3-f)-3\,]}
{(N^{2}+1)(8N^{2}-2Nf-3)} \,, \nn \\
z & = & \frac{y\,\gamma_{0}^{0}}{(\beta_{0}+5\gamma_{m}^{0}-
\gamma_{\sigma}^{0})} \; = \; \frac{-18N^{2}(N^{2}-1)[\,8N^{2}+2N(3-f)-3\,]}
{\pi(N^{2}+1)(8N^{2}-2Nf-3)(10N^{2}+2Nf-15)} \,. \nn
\end{eqnarray}
The element (1,4) of the matrix $\hat R$ which describes the mixing
between $(g_{s}\bar q\sigma Fq)$ and $m^{5}$ is new. We should remark
that strictly speaking it is not fully consistent to take the complete
leading order matrix $\hat R$ at the order we have given the coefficient
functions, because some entries of $\hat R$ appear first at
${\cal O}(\alpha_{s})$. Nevertheless we found it convenient to have at
hand the full leading order renormalization group invariant combinations
of the operators $O_{i}$. Use of the leading as well as next-to-leading
order corrections will be made in \cite{jam:92}.

The relation between condensates evaluated at two different scales
is then given by
\begin{equation}
\big<\vec O(Q^{2})\big> \; = \; \hat U(Q^{2},\,\mu_{0}^{2})\,
\big<\vec O(\mu_{0}^{2})\big> \,, \quad {\rm with} \quad
\hat U(Q^{2},\,\mu_{0}^{2}) \; = \; \hat R^{-1}(Q^{2})\,\hat R(\mu_{0}^{2})\,.
\label{eq:8.11}
\end{equation}
The dependence on $m(Q^{2})$ appearing in the evolution matrix
$\hat U(Q^{2},\,\mu_{0}^{2})$ can of course be expressed in terms of
$m(\mu_{0}^{2})$, using the relation
\begin{equation}
m(Q^{2}) \; = \; \left(\frac{\al(\mu_{0}^{2})}{\al(Q^{2})}\right)^{d_{m}^{\,0}}
m(\mu_{0}^{2}) \,,
\label{eq:8.12}
\end{equation}
such that the $Q^{2}$ dependence of the evolution matrix enters
only through $\al(Q^{2})$.

Putting everything together, our final result for the renormalization
group improved coefficient function of eq.~(\ref{eq:2.1}) takes the
form
\begin{eqnarray}
\Pi^{I}(Q^{2}) & = & C_{1}^{I}\big(\mu^{2}=Q^{2}\big) + \frac{\pi h_{0}^{I}}
{\beta_{0}}\left[\,\frac{1}{\al(Q^{2})}-\frac{1}{\al(\mu^{2})}\,\right] \nn\\
\smvs
& + & \vec C^{I^{T}}\!\big(\mu^{2}=Q^{2}\big)\,\hat U(Q^{2},\,\mu_{0}^{2})\,
\big<\vec O(\mu_{0}^{2})\big> \,,
\label{eq:8.13}
\end{eqnarray}
where the appropriate powers of $1/q^{2}$ have been absorbed into
$\vec C^{I}$.

\pagebreak
\newsection{Summary}

We have calculated the coefficient functions of the quark and the
mixed quark gluon condensate for scalar, pseudoscalar, vector, and
axialvector current correlators to all orders in the quark masses,
in the framework of the operator product expansion. For completeness
the coefficient functions for the unit operator as well as the gluon
condensate are reviewed.

The proper factorization of short and long distance contributions
has been performed, which requires renormalization of the condensates
to the order considered. It is found that it is only possible to
absorb all long distance contributions into the condensates, if the OPE
is expressed in terms of {\em non}-normal-ordered operators. This fact
necessitates the inclusion of mixing with the unit operator under
renormalization.

The resulting coefficient functions were improved with the help of the
renormalization group equation. This is trivial in the case of the
unit operator, but is somewhat more complicated for higher dimensional
operators, since mixing under renormalization has to be taken into
account. The scale invariant combinations up to operators of dimension
5, which are needed for the renormalization group improvement of the
condensates, are given. The inclusion of mixing with the unit operator
in the scale invariant of dimension 5 is new.

An application of our results to an improved determination of the current
strange quark mass, as well as a discussion of higher order $\alpha_{s}$
corrections will be presented in a forthcoming publication \cite{jam:92}

\vskip 1cm \noindent
{\Large\bf Acknowledgement}

\noindent
We would like to thank A. J. Buras, K. G. Chetyrkin, and P. H. Weisz
for helpful discussions. The Feynman diagrams were drawn with the
aid of the program {\em feynd}, written by S. Herrlich.

\newpage
\noindent

\appendix{\LARGE\bf Appendices}
\newsection{Expansions in one small mass $m$}

Here we list the expansions of the coefficient functions $\Cbar_{i}$ in
$m$, up to operators of dimension 6. Since the result in the case of
$\Cbar_{1}$ is very lengthy, it will only be presented up to dimension 4.
With $W=M^{2}-q^{2}$, we find for the

Scalar and Pseudoscalar currents:
\begin{eqnarray}
\Cbar_{1}^{\,S,P} & = & \frac{N}{8\pi^{2}}\Biggl\{2q^{2}-3M^{2}+\frac{M^{4}}
{q^{2}}\ln\frac{M^{2}}{W}-(2M^{2}-q^{2})\ln\frac{\mu^{2}}{W} \nn\\
& & \mp\,2mM\biggl[2-\frac{M^{2}}{q^{2}}\ln\frac{M^{2}}{W}+\ln\frac{\mu^{2}}
{W}\biggr]-2m^{2}\biggl[1+\ln\frac{\mu^{2}}{W}\biggr] \nn\\
& & \pm\,2\frac{m^{3}M}{W}\biggl[1-\frac{M^{2}}{q^{2}}\ln\frac{M^{2}}{W}-
\ln\frac{m^{2}}{W}\biggr]+\frac{m^{4}}{W^{2}}\biggl[M^{2}-\frac{3}{2}q^{2}\nn\\
& & -\,\frac{M^{4}}{q^{2}}\ln\frac{M^{2}}{W}-(2M^{2}-q^{2})\ln\frac{m^{2}}{W}
\biggr]\Biggr\}+{\cal O}(m^{5}) \\
\Cbar_{\bar qq}^{\,S,P} & = & \pm\,\frac{Mq^{2}}{W}+\frac{mq^{2}(2M^{2}-
q^{2})}{2W^{2}}\pm\frac{m^{2}M^{3}q^{2}}{W^{3}}+\frac{m^{3}M^{4}q^{2}}
{W^{4}}+{\cal O}(m^{4}) \\
\Cbar_{\bar QQ}^{\,S,P} & = & -\,\frac{M}{2}\mp m\pm
\frac{m^{3}}{W}+{\cal O}(m^{4}) \,, \qquad \vert q^{2}\vert>M^{2} \\
\Cbar_{FF}^{\,S,P} & = &\frac{1}{12W}\Biggl\{\mp\,\frac{Mq^{2}}{m}-\frac{q^{4}}
{2W}\mp\frac{mq^{4}}{MW^{2}}\biggl[q^{2}+6M^{2}+6M^{2}\ln\frac{mM}{W}
\biggr] \nn\\
& & -\,\frac{m^{2}q^{4}}{W^{3}}\biggl[q^{2}+7M^{2}+6M^{2}\ln\frac{mM}{W}
\biggr]\Biggr\}+{\cal O}(m^{3}) \\
\Cbar_{\bar qFq}^{\,S,P} & = & \mp\,\frac{Mq^{6}}{2W^{3}}-\frac{mM^{2}q^{6}}
{2W^{4}}+{\cal O}(m^{2}) \\
\Cbar_{\bar QFQ}^{\,S,P} & = & \mp\,\frac{mq^{2}}{2W}+{\cal O}(m^{2}) \,,
\qquad \vert q^{2}\vert>M^{2}
\end{eqnarray}

Vector and Axialvector currents:
\begin{eqnarray}
\Cbar_{1}^{\,V,A} & = & \frac{N}{24\pi^{2}}\Biggl\{\frac{10}{3}q^{2}+4M^{2}-
4\frac{M^{4}}{q^{2}}+2(2M^{2}-3q^{2})\frac{M^{4}}{q^{4}}\ln\frac{M^{2}}{W}+
2q^{2}\ln\frac{\mu^{2}}{W} \nn\\
& & +\,6m^{2}\biggl[1+2\frac{M^{2}}{q^{2}}-2\frac{M^{4}}{q^{4}}\ln\frac{M^{2}}
{W}\biggr]-3\frac{m^{4}}{W^{2}}\biggl[\frac{(2M^{2}-q^{2})^{2}}{q^{2}} \nn\\
& & -\,2(2M^{2}-3q^{2})\frac{M^{4}}{q^{4}}\ln\frac{M^{2}}{W}+2q^{2}
\ln\frac{m^{2}}{W}\biggr]\Biggr\}+{\cal O}(m^{5}) \\
\Cbar_{\bar qq}^{\,V,A} & = & \frac{mq^{4}}{W^{2}}+\frac{2m^{3}q^{4}
(4M^{2}-q^{2})}{3W^{4}}+{\cal O}(m^{4}) \\
\Cbar_{\bar QQ}^{\,V,A} & = & M-\frac{2M^{3}}{3q^{2}}+
\frac{2m^{2}M}{q^{2}}+{\cal O}(m^{4}) \,, \qquad \vert q^{2}\vert>M^{2} \\
\Cbar_{FF}^{\,V,A} & = & \frac{-\,q^{4}}{12W^{2}}\Biggl\{1+\frac{2m^{2}}
{W^{2}}\biggl[q^{2}+7M^{2}+6M^{2}\ln\frac{mM}{W}\biggr]\Biggr\}+
{\cal O}(m^{3}) \\
\Cbar_{\bar qFq}^{\,V,A} & = & -\,\frac{mM^{2}q^{6}}{W^{4}}+{\cal O}(m^{2}) \\
\Cbar_{\bar QFQ}^{\,V,A} & = & \mp\,\frac{2mM^{2}}{3W}+{\cal O}(m^{2}) \,,
\qquad \vert q^{2}\vert>M^{2}
\end{eqnarray}

\newsection{The derivation of eq.~(6.4)}

We consider the non-local vacuum expectation value
$\lnp g_{s}\bar q_{\alpha}(0)F_{\mu\nu}(0)\,q_{\beta}(x)\rnp$,
from which we want to project out the local mixed condensate
$\lnp g_{s}\bar q_{\alpha}(0)F_{\mu\nu}(0)\,q_{\beta}(0)\rnp$.
Taylor expansion of the quark field yields
\begin{equation}
\lnp g_{s}\bar q_{\alpha}(0)F_{\mu\nu}(0)\,q_{\beta}(x)\rnp \; = \;
\sum_{k=0}^{\infty}\frac{1}{k!}\,x^{\mu_{1}}\!\ldots x^{\mu_{k}}\,
\lnp g_{s}\bar q_{\alpha}(0)F_{\mu\nu}(0)D_{\mu_{1}}\ldots D_{\mu_{k}}
q_{\beta}(0)\rnp \,,
\label{eq:B.1}
\end{equation}
where we have used the properties of the fixed-point gauge to express
the ordinary derivatives in terms of covariant derivatives
(see \cite{pas:84} and refs. therein). However, since we are not
interested in the contributions from higher dimensional operators,
we are free to replace the covariant derivatives by the ordinary ones,
and work with the free equation of motion $(i\!\Slash\partial-m)q(x)=0$.

An Ansatz for $\lnp g_{s}\bar q_{\alpha}(0)F_{\mu\nu}(0)\,\partial_{\mu_{1}}
\ldots \partial_{\mu_{k}}q_{\beta}(0)\rnp$ has to satisfy two conditions
which come from the Lorentz-structure of eq.~(\ref{eq:B.1}): it has to be
antisymmetric in $\mu$ and $\nu$, but totally symmetric in
$\mu_{1}\ldots\mu_{k}$, due to contraction with the totally symmetric
tensor $x^{\mu_{1}}\!\ldots x^{\mu_{k}}$.

Let us split the sum in (\ref{eq:B.1}) into ``even'' and ``odd'' terms,
in order to simplify our notation,
\begin{eqnarray}
\lnp g_{s}\bar q_{\alpha}F_{\mu\nu}\,q_{\beta}(x)\rnp & = &
\sum_{n=0}^{\infty}\frac{1}{(2n)!}\,x^{\mu_{1}}\!\ldots x^{\mu_{2n}}\,
\lnp g_{s}\bar q_{\alpha}F_{\mu\nu}\,\partial_{\mu_{1}}\ldots
\partial_{\mu_{2n}} q_{\beta}\rnp \label{eq:B.2} \\
\smvs
& + & \sum_{n=1}^{\infty}\frac{1}{(2n-1)!}\,x^{\mu_{1}}\!\ldots x^{\mu_{2n-1}}
\,\lnp g_{s}\bar q_{\alpha}F_{\mu\nu}\,\partial_{\mu_{1}}\ldots
\partial_{\mu_{2n-1}} q_{\beta}\rnp \,, \nn
\end{eqnarray}
where we have only explicitly kept the $x$-dependence.

Consider first the even terms, $k=2n$. There are two possibilities to
construct Lorentz-invariant tensors which fulfill the above requirements.
Therefore, our Ansatz has the following general form
\begin{eqnarray}
\lnp g_{s}\bar q_{\alpha}F_{\mu\nu}\,\partial_{\mu_{1}}\ldots
\partial_{\mu_{2n}} q_{\beta}(0)\rnp & = &
A_{2n}\big(\sigma_{\mu\nu}\big)_{\beta\alpha}S_{\mu_{1}\!\ldots\mu_{2n}}
\label{eq:B.3} \\
\smvs & & \hspace{-2cm}
+\,B_{2n}\sum_{i=1}^{2n}\Big\{\,\big(\sigma_{\mu\mu_{i}}\big)_{\beta
\alpha}S_{\mu_{1}\!\ldots\hat\mu_{i}\ldots\mu_{2n}\nu}+\big(\sigma_{\mu_{i}
\nu}\big)_{\beta\alpha}S_{\mu\mu_{1}\!\ldots\hat\mu_{i}\ldots\mu_{2n}}\,
\Big\} \,. \nn
\end{eqnarray}
Here $S_{\mu_{1}\!\ldots\mu_{2n}}$ denotes the totally symmetric tensor
which consists of a sum of $N_{2n}$ products of $n$ metric tensors, e.g.,
\begin{equation}
S_{\mu_{1}\mu_{2}\mu_{3}\mu_{4}} \; = \; g_{\mu_{1}\mu_{2}}g_{\mu_{3}\mu_{4}}
+g_{\mu_{1}\mu_{3}}g_{\mu_{2}\mu_{4}}+g_{\mu_{1}\mu_{4}}g_{\mu_{2}\mu_{3}} \,.
\label{eq:B.4}
\end{equation}
A hat on an index means, that this index is to be omitted.

By induction it can be seen, that $S_{\mu_{1}\!\ldots\mu_{2n}}$ contains
\begin{equation}
N_{2n} \; = \; \frac{(2n)!}{2^{n}n!}
\label{eq:B.5}
\end{equation}
elements \cite{eli:88}. This number is important for the substitution
of (\ref{eq:B.3}) into (\ref{eq:B.2}), because it appears upon contraction
of the two totally symmetric tensors $S_{\mu_{1}\!\ldots\mu_{2n}}$ and
$x^{\mu_{1}}\!\ldots x^{\mu_{2n}}$,
\begin{equation}
x^{\mu_{1}}\!\ldots x^{\mu_{2n}} S_{\mu_{1}\!\ldots\mu_{2n}} \; = \;
N_{2n} x^{2n} \,.
\label{eq:B.6}
\end{equation}

Contracting eq.~(\ref{eq:B.2}) with
$(\sigma^{\mu\nu})_{\alpha\beta} S^{\mu_{1}\!\ldots\mu_{2n}}$ and
$(\sigma^{\mu\mu_{1}})_{\alpha\beta} S^{\nu\mu_{2}\!\ldots\mu_{2n}}$,
and using the equation of motion, leads to the system of equations
\begin{eqnarray}
D\,A_{2n}+4n\,B_{2n} & = & \frac{\Gamma(D/2)(-m^{2})^{n}}
{4(D-1)2^{n}\Gamma(n+D/2)}\,\lnp g_{s}\bar q\sigma Fq\rnp \,, \nn\\
\smvs
A_{2n}+(D+2n-2)\,B_{2n} & = & 0 \,,
\label{eq:B.7}
\end{eqnarray}
whose solution is
\begin{eqnarray}
A_{2n} & = & \frac{(D+2n-2)\Gamma(D/2)(-m^{2})^{n}}{8(D-1)(D-2)
2^{n}\Gamma(n+1+D/2)}\,\lnp g_{s}\bar q\sigma Fq\rnp \,, \nn \\
\smvs
B_{2n} & = & \frac{-\,\Gamma(D/2)(-m^{2})^{n}}{8(D-1)(D-2)2^{n}
\Gamma(n+1+D/2)}\,\lnp g_{s}\bar q\sigma Fq\rnp \,, \quad n\geq 1
\label{eq:B.8} \\ \smvs
B_{0} & = & 0 \,. \nn
\end{eqnarray}

Let us now come to the odd terms of eq.~(\ref{eq:B.2}), $k=2n-1$.
Here the symmetry considerations allow for the following Ansatz,
\begin{eqnarray}
\lnp g_{s}\bar q_{\alpha}F_{\mu\nu}\,\partial_{\mu_{1}}\ldots
\partial_{\mu_{2n-1}} q_{\beta}(0)\rnp & = &
A_{2n-1}\big(\sigma_{\mu\nu}S_{\mu_{1}\!\ldots\mu_{2n-1}}\big)_{\beta\alpha}
\label{eq:B.9} \\
\smvs & & \hspace{-2cm}
+\,B_{2n-1}\,\Big\{\,\big(\gamma_{\mu}S_{\mu_{1}\!\ldots\mu_{2n-1}\nu}
\big)_{\beta\alpha}-\big(\gamma_{\nu}S_{\mu\mu_{1}\!\ldots\mu_{2n-1}}
\big)_{\beta\alpha}\,\Big\} \,. \nn
\end{eqnarray}
The tensor $S_{\mu_{1}\!\ldots\mu_{2n-1}}$ is the sum of $N_{2n-1}$
products of one $\gamma$-matrix with $n-1$ metric tensors, where
\begin{equation}
N_{2n-1} \; = \; \frac{(2n-1)!}{2^{n-1}(n-1)!} \,.
\label{eq:B.10}
\end{equation}
As an example, for $n=2$
\begin{equation}
S_{\mu_{1}\mu_{2}\mu_{3}} \; = \; \gamma_{\mu_{1}}g_{\mu_{2}\mu_{3}}+
\gamma_{\mu_{2}}g_{\mu_{1}\mu_{3}}+\gamma_{\mu_{3}}g_{\mu_{1}\mu_{2}} \,.
\label{eq:B.11}
\end{equation}
Similar contractions as in the even case lead to
\begin{eqnarray}
(D-4)\,A_{2n-1}+2i\,B_{2n-1} & = & \frac{i\,\Gamma(D/2)(-m^{2})^{n}}
{4m(D-1)2^{n}\Gamma(n+D/2)}\,\lnp g_{s}\bar q\sigma Fq\rnp \,, \nn\\
\smvs
A_{2n-1}-i\,B_{2n-1} & = & 0 \,,
\label{eq:B.12}
\end{eqnarray}
so that
\begin{eqnarray}
A_{2n-1} & = & \frac{i\,\Gamma(D/2)(-m^{2})^{n}}{4m(D-1)(D-2)
2^{n}\Gamma(n+D/2)}\,\lnp g_{s}\bar q\sigma Fq\rnp \,, \nn \\
\smvs
B_{2n-1} & = & \frac{\Gamma(D/2)(-m^{2})^{n}}{4m(D-1)(D-2)
2^{n}\Gamma(n+D/2)}\,\lnp g_{s}\bar q\sigma Fq\rnp \,.
\label{eq:B.13}
\end{eqnarray}

Putting now things together, i.e., substituting (\ref{eq:B.3}),
(\ref{eq:B.8}), (\ref{eq:B.9}), and (\ref{eq:B.13}) into
(\ref{eq:B.2}), and carrying out the contractions of
$x^{\mu_{1}}\!\ldots x^{\mu_{n}}$ with the corresponding
symmetric tensors, we arrive after some algebra at the simple
result of eq.~(\ref{eq:6.3})

\newpage


\end{document}